\documentclass[conference]{IEEEtran}
\IEEEoverridecommandlockouts
\pdfoutput=1 
\usepackage{cite}
\usepackage{amsmath,amssymb,amsfonts}
\usepackage{algorithmic}
\usepackage{graphicx}

\usepackage{textcomp}
\usepackage{xcolor}
\usepackage{makecell}

\def\BibTeX{{\rm B\kern-.05em{\sc i\kern-.025em b}\kern-.08em
    T\kern-.1667em\lower.7ex\hbox{E}\kern-.125emX}}

\begin{document}
	
\bibliographystyle{IEEEtran}


\title{Dynamic Sub-array Based Modeling for Large-Scale RIS-assisted mmWave UAV Channels}

\author{
\IEEEauthorblockN{Baiping Xiong\IEEEauthorrefmark{1}\IEEEauthorrefmark{2}, Zaichen Zhang\IEEEauthorrefmark{1}\IEEEauthorrefmark{2}, and Jiangzhou Wang\IEEEauthorrefmark{3} }  

\IEEEauthorblockA{\IEEEauthorrefmark{1}National Mobile Communications Research Laboratory, Southeast University, Nanjing 210096, China}  
\IEEEauthorblockA{\IEEEauthorrefmark{2}Purple Mountain Laboratories, Nanjing 211111, China}  
\IEEEauthorblockA{\IEEEauthorrefmark{3}School of Engineering, University of Kent, Canterbury CT2 7NT, U.K.}  
\IEEEauthorblockA{Emails: xiongbp@seu.edu.cn, zczhang@seu.edu.cn, j.z.wang@kent.ac.uk.}  
}

\maketitle

\begin{abstract}
Large-scale reconfigurable intelligent surface (RIS) can effectively enhance the performance of millimeter wave (mmWave) unmanned aerial vehicle (UAV) to ground communication link with obstructed line-of-sight (LoS) path by exploiting more reflecting units. However, the non-negligible array dimension of large-scale RIS and the mobile property of the terminals bring significant variations in propagation characteristics, making conventional channel models inapplicable. To address this issue, we propose a dynamic sub-array partition scheme to divide the large-scale RIS into sub-arrays by exploiting the Rayleigh distance criterion and the mobile property of the transceivers. Based on the proposed scheme, a novel non-stationary channel model for large-scale RIS auxiliary mmWave UAV-to-ground mobile networks is developed, which outperforms existing models with well balance between model complexity and accuracy. Numerical results are provided to verify our analysis.
\end{abstract}


\section{Introduction}

With the evolution of meta-materials and wireless communications, an interdisciplinary concept called reconfigurable intelligent surface (RIS) has been proposed, which provides new degrees of freedom for converting wireless channels from uncontrollable to partially controllable, building a smart propagation environment for communications \cite{Basar_future_RIS}. Specifically, the RIS technology has been considered a cost- and energy-efficient manner to boost the millimeter wave (mmWave) unmanned aerial vehicle (UAV) to ground communication link under direct path easy blocking scenarios, such as city environments with dense buildings, by contributing a virtual line-of-sight (V-LoS) path between the UAV transmitter and ground mobile receiver (MR) \cite{JY_nonTer}. However, the product-fading attenuation effect experienced by RIS has heavily limited its applications to support the performance requirements of next generation wireless networks \cite{Double_fading}.

For the purpose of enhancing the performance of the RIS auxiliary link to a reasonable level, one of the candidate solutions is to replace conventional passive RIS with active RIS by integrating reflection-type amplifier to each RIS unit, with which the incident signals will be amplified by utilizing the energy converted from DC power supply \cite{active_amplify_1}, \cite{active_amplify_2}. Specifically, the authors in \cite{Double_fading} proposed a novel active RIS architecture and examined it via experimental measurements, where the results from the capacity gain analysis showed that active RIS has the potential to overcome the product-fading attenuation effect experienced by passive RIS. The authors in \cite{ActiveRISTHz} considered the sum rate maximization of an active RIS-assisted terahertz (THz) communication system and showed that the proposed scheme is superior to passive RIS as well as multi-antenna amplify-and-forward (AF) relaying. Despite its attractive advantages, active RIS still has a long way to go and may not be a top priority at present. This is mainly due to that although active RIS reflects signals with amplification, it suffers from a large amount of extra power consumption as compared to passive RIS. Meanwhile, the fabrication of additional active amplifiers will also increase the cost of system design. In addition, along with the desired signal components, active RIS will unintentionally amplify all the incident waves, thus bringing extra interference and noise to the receiver.

Another roadmap to this challenge follows the evolution route of multi-antenna technology, i.e., multiple-input multiple-output (MIMO) $\to$ massive MIMO $\to$ ultra massive MIMO $\to$ extremely large (massive) MIMO \cite{ExLMIMO}, etc., whose feasibility stems from the square-law power gain of RIS \cite{RISsquarelaw}. The use of higher frequency bands, e.g., mmWave, THz, and/or optical, makes it possible to fabricate more units within finite array dimension; and meanwhile the low-cost property of passive RIS enables its large-scale deployment. Integrating more reflecting units will contribute higher power gain, but also brings non-negligible array dimension. This will cause larger array Rayleigh distance and makes far-field propagation condition invalid. Moreover, the mobile property of the transceivers (UAV and MR) enables the variation of RIS-transceiver distances, which will cause the transition of propagation condition from near-field to far-field or vice versa. In this case, the existing planar wavefront assumption based models for far-field propagation condition, such as \cite{RISsquarelaw}-\cite{Basar_planar}, no longer applicable. On the other hand, spherical wavefront assumption based models obtain excellent accuracy but suffers from high computation complexity.

Following the intentions of low-cost and low-energy consumption of RIS, this paper considers a large-scale RIS auxiliary mmWave aerial to ground communication system, in which the large-scale RIS is coated on the surface of a building as depicted in Fig. 1. To address the aforementioned issues experienced by large-scale RIS, we propose a soft-level dynamic sub-array partition scheme to partition the large-scale RIS into sub-arrays by exploiting the Rayleigh distance criterion as well as the mobile property of the transceivers. Subsequently, far-field propagation condition is satisfied by the sub-arrays and hence planar-wavefront assumption can be adopted to these sub-arrays. Based on the proposed scheme, we derive the complex channel response of the proposed channel model, which is verified to be superior to existing models with well balance between modeling complexity and accuracy.

\emph{Notation:} Throughout this paper, non-boldface, boldface lowercase, and boldface uppercase letters denote scalar, vector, and matrix, respectively; $\vert \cdot \vert$, $(\cdot)^{\emph{T}}$, $(\cdot)^{\ast}$, and $\langle\cdot , \cdot\rangle$ stand for absolute value, transpose, complex conjugate, and vector dot product, respectively; $\mathbb{E}\{\cdot\}$, $\min\{\cdot, \cdot\}$, ${\lfloor \cdot \rceil}$, and $\text{mod}\{\cdot, \cdot\}$ take the expectation, minimum value, downward integer, and modulo division, respectively.

\begin{figure}[!t]  
\centerline{\includegraphics[width=2.75in,height=1.9in]{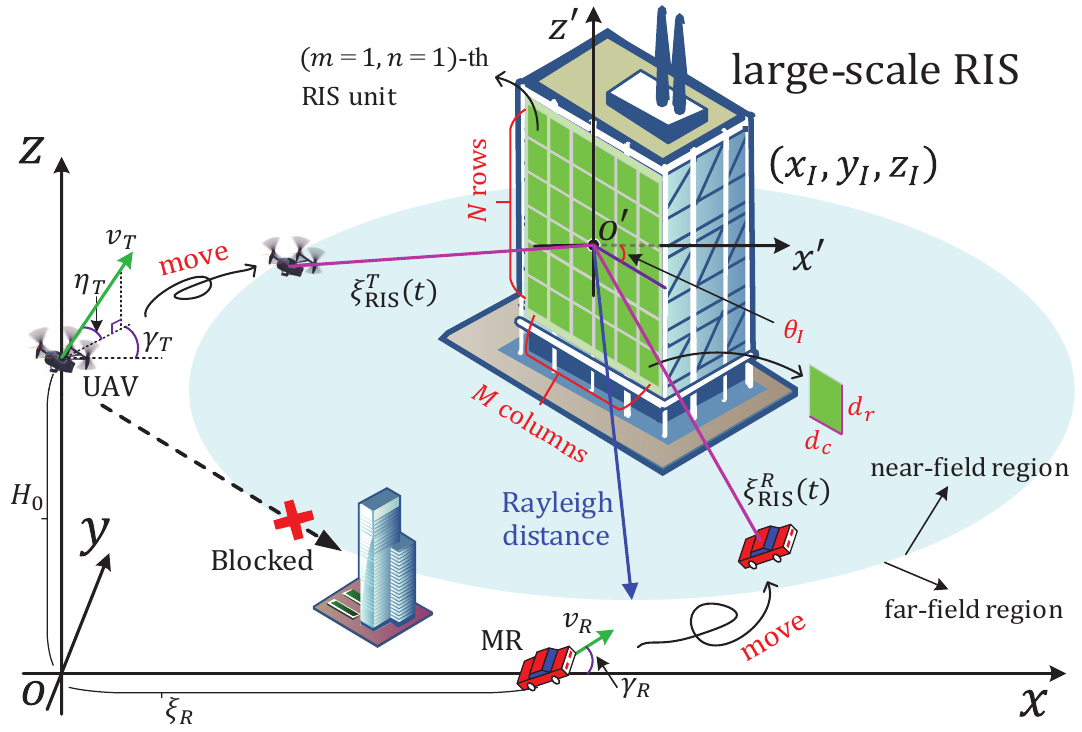}}
\caption{A physical illustration of the proposed large-scale RIS assisted mmWave UAV communications with dynamic sub-array partition.}
\end{figure}

\section{System Model}

As depicted in Fig. 1, we consider a wireless communication system which suffers from severe performance deterioration due to the blocking line-of-sight (LoS) path between a mobile unmanned aerial vehicle (UAV) with velocity vector $\textbf{v}_T$ and a moving MR with velocity vector $\textbf{v}_R$. The velocity vectors are expressed as $\textbf{v}_T = v_T [\cos\eta_T\cos\gamma_T, \cos\eta_T\sin\gamma_T, \sin\eta_T]^{\emph{T}}$ and $\textbf{v}_R = v_R [\cos\gamma_R, \sin\gamma_R, 0]^{\emph{T}}$, respectively. To enhance the communication performance, a large-scale reconfigurable intelligent surface (RIS) is deployed on the facade of a building between the UAV and MR, which contributes a virtual-LoS (VLoS) link between the transceivers. Omnidirectional uniform linear arrays (ULAs) with antenna numbers $M_T$ and $M_R$, as well as antenna element spacings $\delta_T$ and $\delta_R$, are employed at the UAV and MR, respectively. The large-scale RIS is composed of $M$ columns and $N$ rows, with unit size denoted by $d_c$ and $d_r$ in the column and row directions, respectively. Therefore, the total number of reflecting units is $MN$. Typically, the unit size $d_{c(r)}$ is on the sub-wavelength scale (e.g., $\lambda/10 \sim \lambda/2$), where $\lambda$ is the wavelength, while the gap between adjacent unit is ignored. Following the coordinate system definition in \cite{X_ReLoS}, the locations of the central points of UAV's ULA, large-scale RIS, and MR's ULA at the initial instant are denoted by ($0, 0, H_0$), ($x_I, y_I, z_I$), and ($\xi_R, 0, 0$), respectively. To enhance the model generality, the orientation angles of the antenna arrays are considered. Specifically, the azimuth and elevation orientation angles of the UAV's ULA are denoted by $\psi_T$ and $\phi_T$, respectively, whereas at the MR, they are denoted by $\psi_R$ and $\phi_R$, respectively. The large-scale RIS is located on the facade of a building, thus only a horizontal rotation angle of $\theta_I$ is considered. The definitions of the key model parameters are summarized in Table I.

\begin{table}
\footnotesize    
\centering
\caption{Summary of Key Parameters Definitions}
\begin{tabular}{|c|c|}
\hline
$\xi^{T}_{\text{RIS}}(t)$, $\xi^{R}_{\text{RIS}}(t)$  &  distances between UAV/MR and center of $\text{RIS}$   \\
\hline
$\xi^{T}_{m n}(t)$, $\xi^{R}_{m n}(t)$  &  distances between UAV/MR and ($m, n$)-th RIS unit   \\
\hline
$\alpha^{T}_{mn}(t)$, $\beta^{T}_{mn}(t)$  &  AAoD and EAoD from UAV to ($m, n$)-th RIS unit      \\
\hline
$\alpha^{R}_{mn}(t)$, $\beta^{R}_{mn}(t)$  &  AAoA and EAoA from ($m, n$)-th RIS unit to MR     \\
\hline
$\chi_{mn}(t)$  &  reflection amplitude of ($m, n$)-th RIS unit  \\
\hline
$\varphi_{mn}(t)$  &  reflection phase of ($m, n$)-th RIS unit   \\
\hline
$v_T$, $\gamma_T$, $\eta_T$  & motion speed, azimuth/elevation directions of UAV  \\
\hline
$v_R$, $\gamma_R$  &  motion speed and azimuth direction of MR    \\
\hline
\end{tabular}
\end{table}

In this paper, we denote $s$ as the transmitted symbol with power $\mathbb{E}\{ss^\ast \} = P_T$, $\textbf{H}_{\text{TR}}(t, f) \in \mathbb{C}^{M_R \times M_T}$ as the frequency domain end-to-end channel matrix between the transceiver, and then the complex baseband received signal model can be expressed as \cite{X_ReLoS}, \cite{BeamCIR}, 
\begin{eqnarray}
y(t) \hspace*{-0.225cm}&=&\hspace*{-0.225cm}   \textbf{f}^{\emph{T}}_{\text{MR}}(t) \textbf{H}_{\text{TR}}(t, f) \textbf{f}_{\text{UAV}}(t) s +  n(t)    , 
\end{eqnarray}
where $\textbf{f}_{\text{UAV}}(t) \in \mathbb{C}^{M_T \times 1}$ and $\textbf{f}_{\text{MR}}(t) \in \mathbb{C}^{M_R \times 1}$ denote the UAV transmit beamforming and MR receive combining vectors, respectively, and $n(t)$ is the zero-mean complex Gaussian noise with variance $\sigma^2$. Once the MIMO channel matrix $\textbf{H}_{\text{TR}}(t, f)$ is derived, the beamforming/combining vector $\textbf{f}_{\text{UAV}/\text{MR}}(t)$ can be obtained via the maximal-ratio combining strategy \cite{Goldsmith}. By taking inverse Fourier transformation, the frequency domain channel matrix $\textbf{H}_{\text{TR}}(t, f)$ can be converted to its time domain counterpart, i.e., $\textbf{H}_{\text{TR}} (t, \tau) = \int \textbf{H}_{\text{TR}}(t, f) e^{j 2\pi f \tau} d f  = \big[ h_{pq}(t, \tau) \big]_{M_R \times M_T}$, where $h_{pq}(t, \tau)$ denotes the complex channel impulse response (CIR) including path loss between the $p$-th ($p = 1, 2, ..., M_T$) UAV antenna and the $q$-th ($q = 1, 2, ..., M_R$) MR antenna.

Benefit from the directional transmission brought by beamforming, it is reasonable to assume that the UAV transmitter concentrates the majority of the signal power on the MR through the reflection of large-scale RIS. Therefore, we ignore the components independent of larger-scale RIS in the following discussions due to their negligible power gains. Subsequently, we can present a general form of $h_{pq}(t, \tau)$ as
\allowdisplaybreaks[4]
\begin{eqnarray}
h_{pq}(t, \tau)   && \nonumber \\ [0.15cm] 
&&\hspace*{-1.825cm}= \sqrt{ \frac{ \Omega_{\text{RIS}}(t) }{ \Upsilon_{pq}(t) } }  \sum^{N}_{n = 1} \sum^{M}_{m = 1}   \chi_{mn}(t)  e^{j \varphi_{mn}(t) }  \nonumber \\ [0.15cm]
&&\hspace*{-1.6cm}\times  e^{- j \frac{2\pi}{\lambda} \big( \xi^{T}_{mn}(t) + \xi^{R}_{mn}(t) \big)  }    \times  e^{ j \frac{2\pi}{\lambda} \langle \emph{\textbf{e}}^{T}_{mn}(t), \;  \textbf{d}^{T}_{p}  \rangle }    e^{ j \frac{2\pi}{\lambda} \langle \emph{\textbf{e}}^{R}_{mn}(t), \;  \textbf{d}^{R}_{q}  \rangle }   \nonumber \\ [0.15cm] 
&&\hspace*{-1.6cm}\times  e^{ j \frac{2\pi}{\lambda} \langle  \textbf{v}_T t, \;  \emph{\textbf{e}}^{T}_{mn}(t)  \rangle }  \times  e^{ j \frac{2\pi}{\lambda} \langle \textbf{v}_R t, \;  \emph{\textbf{e}}^{R}_{mn}(t)  \rangle }     \delta\big(\tau - \tau_{mn}(t)\big)  ,
\end{eqnarray}
where $\Omega_{\text{RIS}}(t)$ denotes the end-to-end path power gain including the path loss between the UAV and MR, $\Upsilon_{pq}(t)$ is the normalized factor, and $\tau_{mn}(t) = \big(\xi^{T}_{mn}(t) + \xi^{R}_{mn}(t)\big)/c$ is the path delay with $c = 3.0 \times 10^8$ m/s. The $\textbf{d}^{T}_{p}$ and $\textbf{d}^{R}_{q}$ are the distance vectors from the centers of the UAV's and MR's ULAs to the $p$-th transmit and $q$-th receive antennas, respectively, which can be expressed as
\begin{eqnarray}
\textbf{d}^{T/R}_{i}  \hspace*{-0.225cm}&=&\hspace*{-0.225cm}   \frac{M_{T/R} - 2 i + 1}{2} \delta_{T/R}
\begin{bmatrix}
\hspace*{0.05cm}  \cos\phi_{T/R} \cos\psi_{T/R}  \hspace*{0.05cm}  \\[0.05cm]
\hspace*{0.05cm}  \cos\phi_{T/R} \sin\psi_{T/R}   \hspace*{0.05cm}   \\[0.05cm]
\hspace*{0.05cm}  \sin\phi_{T/R}   \hspace*{0.05cm}
\end{bmatrix} ,
\end{eqnarray}
where $i = p$ for transmit antenna, e.g., $\textbf{d}^{T}_{p}$, and $i = q$ for receive antenna, e.g., $\textbf{d}^{R}_{q}$, respectively. Moreover, $\emph{\textbf{e}}^{T}_{mn}(t)$ and $\emph{\textbf{e}}^{R}_{mn}(t)$ denote the unit directional vectors from UAV and MR to the ($m, n$)-th ($m = 1, 2, ..., M; n = 1, 2, ..., N$) reflecting unit, respectively, which are expressed as
\begin{eqnarray}
\emph{\textbf{e}}^{T/R}_{mn}(t)  \hspace*{-0.225cm}&=&\hspace*{-0.225cm} 
\begin{bmatrix}
\hspace*{0.05cm}  \cos\beta^{T/R}_{mn}(t) \cos\alpha^{T/R}_{mn}(t)  \hspace*{0.05cm}  \\[0.05cm]
\hspace*{0.05cm}  \cos\beta^{T/R}_{mn}(t) \sin\alpha^{T/R}_{mn}(t)   \hspace*{0.05cm}   \\[0.05cm]
\hspace*{0.05cm}  \sin\beta^{T/R}_{mn}(t)   \hspace*{0.05cm}
\end{bmatrix} .
\end{eqnarray}

We have to mention that (2) provides a general description of the CIR $h_{pq}(t, \tau)$ based on the spherical wavefront model, which attains excellent accuracy but suffering from very high complexity, especially for large-scale RIS mobile scenarios. Planar wavefront model-based description of the CIR $h_{pq}(t, \tau)$ has been widely used in literature because of its low complexity, but cannot meet with the accuracy requirement in near-field propagation condition. The large-scale RIS assisted mobile networks show new characteristics as described in Fig. 1. On the one hand, a large-scale RIS has non-negligible array dimension, which causes considerable Rayleigh distance as compared to the distances between RIS and the terminals, resulting in near-field propagation condition. In this case, the conventional planar wavefront model adopted for far-field propagation condition is no longer valid due to insufficient accuracy. On the other hand, owing to the motion of the terminals, the distances between RIS and the terminals are time-varying. This may cause the propagation condition transitions from near-field to far-field or vice versa. In this case, the spherical wavefront model suffers from high computation complexity. To address the aforementioned issue, we will develop a dynamic sub-array partition scheme for large-scale RIS assisted mobile communications, which achieves a well balance between modeling accuracy and complexity.

\section{Proposed Dynamic Sub-array Partition Scheme}

Rayleigh distance is widely used for characterizing the boundary between near-field and far-field propagations \cite{Raydis}. When the terminal is outside of the Rayleigh distance range of the large-scale RIS, the wave propagation is far-field and hence planar wavefront signal model can be adopted, in which different reflecting elements share the same signal angle. Otherwise, the wave propagation is near-field and spherical wavefront signal model should be applied. The proposed dynamic sub-array partition scheme achieves a balance between modeling complexity and accuracy by exploiting this characteristic as well as the mobile property of the transceivers. Specifically, sub-array partition is made to the large-scale RIS so that all the resulting sub-arrays satisfy the far-field propagation condition, thus planar wavefront model can be applied to the sub-arrays. When the terminals move as time goes by, closer to or farther away from the large-scale RIS, the sub-arrays will be re-partitioned to accommodate to the accuracy and complexity requirement of the signal model.

\begin{figure}[!t]  
\centerline{\includegraphics[width=2.3in,height=1.65in]{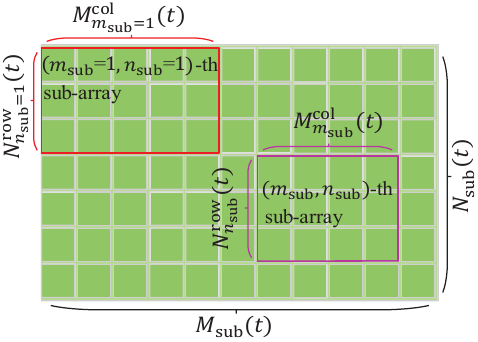}}
\caption{Description of the proposed dynamic sub-array partition scheme.}
\end{figure}

Figure 2 illustrates the sub-array partition result at time instant $t$, where the entire large-scale RIS is evenly partitioned into $M_{\text{sub}}(t) \times N_{\text{sub}}(t)$ sub-arrays. Since we apply no constraints to the total reflecting units numbers $M$ and $N$, the number of reflecting units in different sub-arrays can be different. Specifically, for the ($m_{\text{sub}}, n_{\text{sub}}$)-th ($m_{\text{sub}} = 1, 2, ..., M_{\text{sub}}(t); n_{\text{sub}} = 1, 2, ..., N_{\text{sub}}(t)$) sub-array, the number of reflecting units in the column and row directions are denoted by $M^{\text{col}}_{m_{\text{sub}}} (t)$ and $N^{\text{row}}_{n_{\text{sub}}} (t)$, respectively. In general, the first sub-array, that is, the ($m_{\text{sub}} = 1, n_{\text{sub}} = 1$)-th sub-array, has the most reflecting units among all the sub-arrays. Therefore, if the ($m_{\text{sub}} = 1, n_{\text{sub}} = 1$)-th sub-array meets with the far-field propagation condition, so do all the other sub-arrays. Consequently, the reflecting units number in the ($m_{\text{sub}} = 1, n_{\text{sub}} = 1$)-th sub-array, denoted by $M^{\text{col}}_{m_{\text{sub}} = 1} (t)$ and $N^{\text{row}}_{n_{\text{sub}} = 1} (t)$, respectively, can be determined as 
\begin{eqnarray}
M^{\text{col}}_{m_{\text{sub}} = 1}(t) \hspace*{-0.225cm}&=&\hspace*{-0.225cm} \min \Big\{ \, \bigg\lfloor \,  \sqrt{ \frac{ \lambda \xi_{\text{min}}(t) }{2 d^2_c}} \,  \bigg\rceil, \; M \, \Big\}  , 
\end{eqnarray}  
\begin{eqnarray}
N^{\text{row}}_{n_{\text{sub}} = 1}(t) \hspace*{-0.225cm}&=&\hspace*{-0.225cm}  \min \Big\{ \,  \bigg\lfloor  \,  \sqrt{ \frac{\lambda \xi_{\text{min}}(t) }{2 d^2_r}} \,  \bigg\rceil, \; N \, \Big\}   , 
\end{eqnarray}
where $\xi_{\text{min}}(t)$ denotes the minimum distance from the terminals to the large-scale RIS, and it is calculated by $\xi_{\text{min}}(t) = \min\big\{ \sqrt{ (x_I - r^{T}_{x}(t) )^2 + (y_I - r^{T}_{y}(t) )^2 + (z_I - H_0 - r^{T}_{z}(t) )^2 }$ $, \sqrt{ (x_I - \xi_R - r^{R}_{x}(t) )^2 + (y_I - r^{R}_{y}(t) )^2 + z^{2}_{I} } \big\}$. For the UAV transmitter, $r^{T}_{x}(t) = v_T t \cos\eta_T\cos\gamma_T$, $r^{T}_{y}(t) = v_T t$ $\cos\eta_T\sin\gamma_T$, and $r^{T}_{z}(t) = v_T t \sin\eta_T$, respectively; whereas at the receiver, we have $r^{R}_{x}(t) = v_R t \cos\gamma_R$ and $r^{R}_{y}(t) =$ $v_R t \sin\gamma_R$, respectively. Subsequently, the number of sub-arrays in the column and row directions, they are, $M_{\text{sub}}(t)$ and $N_{\text{sub}}(t)$, respectively, can be further determined as 
\allowdisplaybreaks[4]
\begin{eqnarray}
M_{\text{sub}} (t)  =
\begin{cases} 
\frac{ M \,-\, \text{mod}\big\{ M, \, M^{\text{col}}_{m_{\text{sub}} = 1}(t) \big\} }{ M^{\text{col}}_{m_{\text{sub}} = 1}(t) } + 1     ,     &\textrm{if $A_1 \neq 0$ }    \\  
\frac{ M }{ M^{\text{col}}_{m_{\text{sub}} = 1}(t) } , &\textrm{if $A_1 = 0$ }  
\end{cases}          ,  
\end{eqnarray}
\begin{eqnarray}
N_{\text{sub}} (t)  =
\begin{cases}   
\frac{ N \,-\, \text{mod}\big\{ N, \, N^{\text{row}}_{n_{\text{sub}} = 1}(t) \big\} }{ N^{\text{row}}_{n_{\text{sub}} = 1}(t) } + 1   ,      &\textrm{if $A_2 \neq 0$ }     \\  
\frac{ N }{ N^{\text{row}}_{n_{\text{sub}} = 1}(t) }  ,    &\textrm{if $A_2 = 0$ }  
\end{cases}           ,
\end{eqnarray}
where $A_1 = \text{mod}\big\{ M, \, M^{\text{col}}_{m_{\text{sub}} = 1}(t) \big\}$ and $A_2 = \text{mod}\big\{ N,$ $\, N^{\text{row}}_{n_{\text{sub}} = 1}(t) \big\}$, respectively. It can be seen that when the sub-array is arbitrarily small, which means that $M^{\text{col}}_{m_{\text{sub}} = 1}(t) = N^{\text{row}}_{n_{\text{sub}} = 1}(t) = 1$, $M_{\text{sub}} (t) = M$, and $N_{\text{sub}} (t) = N$, e.g., each reflecting unit corresponds to one sub-array, the proposed model evolves into the spherical wavefront model with highest complexity. When the sub-array is arbitrarily large, which indicates that $M^{\text{col}}_{m_{\text{sub}} = 1}(t) = M$, $N^{\text{row}}_{n_{\text{sub}} = 1}(t) = N$, and $M_{\text{sub}} (t) = N_{\text{sub}} (t) = 1$, e.g., the entire large-scale RIS corresponds to one sub-array, the proposed model reduces into the planar wavefront model with acceptable accuracy. Overall, in this proposed dynamic sub-array partition scheme, the modeling complexity is reduced by applying planar wavefront signal model to the sub-arrays, while the modeling accuracy is ensured by adopting the dynamic partition strategy taking Rayleigh distance as the metric.

More specifically, we can further derive the number of reflecting units in the column and row directions for the arbitrarily ($m_{\text{sub}}, n_{\text{sub}}$)-th sub-array, they are, $M^{\text{col}}_{m_{\text{sub}}} (t)$ and $N^{\text{row}}_{n_{\text{sub}}} (t)$, respectively, as 
\begin{eqnarray}
M^{\text{col}}_{m_{\text{sub}}}(t)  && \nonumber \\ 
&&\hspace*{-1.975cm}= \hspace*{-0.025cm} 
\begin{cases}  
M^{\text{col}}_{m_{\text{sub}} = 1}(t),  \hspace*{-0.175cm}  &\textrm{if $1  \leq  m_{\text{sub}}  <  M_{\text{sub}}(t)$ }    \\    
M  -  \big( M_{\text{sub}}(t) - 1 \big) M^{\text{sub}}_{m_{\text{sub}} = 1}(t)    ,    \hspace*{-0.175cm}  &\textrm{if  $m_{\text{sub}} = M_{\text{sub}}(t)$ }  
\end{cases}  \hspace*{-0.225cm} ,   \nonumber  \\ 
\end{eqnarray}
\begin{eqnarray}
\hspace*{-0.245cm} N^{\text{row}}_{n_{\text{sub}}}(t)  && \nonumber \\ 
&&\hspace*{-1.775cm}= \hspace*{-0.025cm} 
\begin{cases}  
N^{\text{row}}_{n_{\text{sub}} = 1}(t) ,   \hspace*{-0.175cm} &\textrm{if $1  \leq  n_{\text{sub}}  <  N_{\text{sub}}(t)$ }    \\    
N  -  \big( N_{\text{sub}}(t) - 1 \big) N^{\text{row}}_{n_{\text{sub}} = 1}(t)    ,   \hspace*{-0.175cm}  &\textrm{if  $n_{\text{sub}} = N_{\text{sub}}(t)$ }  
\end{cases}  \hspace*{-0.225cm} .  \nonumber  \\ 
\end{eqnarray}
Subsequently, the distance vector from the origin to the center point of the ($m_{\text{sub}}, n_{\text{sub}}$)-th sub-array, denoted by $\textbf{d}_{(m_{\text{sub}}, n_{\text{sub}})} (t)$, can be expressed as
\begin{eqnarray}
\textbf{d}_{(m_{\text{sub}}, n_{\text{sub}})} (t)  =
\begin{bmatrix}
x_I + A_3 \cos\theta_I    \\ 
y_I + A_3 \sin\theta_I     \\ 
z_I - A_4     
\end{bmatrix}  ,  
\end{eqnarray}
where $A_3 = \frac{1}{2}\big[2(m_{\text{sub}} - 1)M^{\text{col}}_{m_{\text{sub}} = 1}(t) + M^{\text{col}}_{m_{\text{sub}}}(t) - M \big]d_c$ and $A_4 = \frac{1}{2}\big[2(n_{\text{sub}} - 1)N^{\text{col}}_{n_{\text{sub}} = 1}(t) + N^{\text{col}}_{n_{\text{sub}}}(t) - N \big]d_r$, respectively.

Then, following the procedure described in \cite{X_3DRISTCOM}, the time-varying distances from the centers of the UAV's and MR's ULAs to that of the ($m_{\text{sub}}, n_{\text{sub}}$)-th sub-array, denoted by $\xi^{T}_{(m_{\text{sub}}, n_{\text{sub}})}(t)$ and $\xi^{R}_{(m_{\text{sub}}, n_{\text{sub}})}(t)$, respectively, are calculated by
\allowdisplaybreaks[4]
\begin{eqnarray}
\xi^{T}_{(m_{\text{sub}}, n_{\text{sub}})}(t)  \hspace*{-0.225cm}&=&\hspace*{-0.225cm}  \big\{ (x_I + A_3 \cos\theta_I - r^{T}_{x}(t) )^2   \nonumber \\ 
&&\hspace*{-0.225cm}+   (y_I + A_3 \sin\theta_I - r^{T}_{y}(t) )^2  \nonumber \\ 
&&\hspace*{-0.225cm}+   (z_I - A_4 - H_0 - r^{T}_{z}(t) )^2 \big\}^{1/2}  ,  
\end{eqnarray}
\begin{eqnarray}
\xi^{R}_{(m_{\text{sub}}, n_{\text{sub}})}(t)  \hspace*{-0.225cm}&=&\hspace*{-0.225cm}  \big\{ (x_I + A_3 \cos\theta_I - \xi_R - r^{R}_{x}(t) )^2   \nonumber \\ 
&&\hspace*{-0.225cm}+  (y_I + A_3 \sin\theta_I - r^{R}_{y}(t) )^2  \nonumber \\ 
&&\hspace*{-0.225cm}+  (z_I - A_4)^2 \big\}^{1/2}  , 
\end{eqnarray}
and the corresponding time-varying departure angles from UAV to the ($m_{\text{sub}}, n_{\text{sub}}$)-th sub-array, they are, the azimuth departure angle $\alpha^{T}_{(m_{\text{sub}}, n_{\text{sub}})}(t)$ and elevation departure angle $\beta^{T}_{(m_{\text{sub}}, n_{\text{sub}})}(t)$, respectively, are expressed as
\begin{eqnarray}
\beta^{T}_{(m_{\text{sub}}, n_{\text{sub}})}(t)  =  \arcsin \frac{ z_I - A_4 - H_0 - r^{T}_{z}(t) }{ \xi^{T}_{(m_{\text{sub}}, n_{\text{sub}})}(t) },
\end{eqnarray}
\begin{eqnarray}
\hspace*{-0.425cm} \alpha^{T}_{(m_{\text{sub}}, n_{\text{sub}})}(t)  && \nonumber \\  
&&\hspace*{-2.525cm}=  \arccos \frac{ x_I + A_3 \cos\theta_I - r^{T}_{x}(t) }{ \sqrt{ { \xi^{T}_{(m_{\text{sub}}, n_{\text{sub}})} }^2 (t) - (z_I - A_4 - H_0 - r^{T}_{z}(t) )^2 } } ,
\end{eqnarray}
at the receiver, the time-varying arrival angles from the ($m_{\text{sub}}, n_{\text{sub}}$)-th sub-array to the MR, they are, the azimuth arrival angle $\alpha^{R}_{(m_{\text{sub}}, n_{\text{sub}})}(t)$ and elevation arrival angle $\beta^{R}_{(m_{\text{sub}}, n_{\text{sub}})}(t)$, respectively, are expressed as
\begin{eqnarray}
\beta^{R}_{(m_{\text{sub}}, n_{\text{sub}})}(t)  =  \arcsin \frac{ z_I - A_4 }{ \xi^{R}_{(m_{\text{sub}}, n_{\text{sub}})}(t) } ,
\end{eqnarray}
\begin{eqnarray}
\alpha^{R}_{(m_{\text{sub}}, n_{\text{sub}})}(t)  =  \arccos \frac{ x_I + A_3 \cos\theta_I - \xi_R - r^{R}_{x}(t) }{ \sqrt{ { \xi^{R}_{(m_{\text{sub}}, n_{\text{sub}})} }^2 (t) - (z_I - A_4)^2 } }. 
\end{eqnarray}

Finally, with the aforementioned derivations, the CIR $h_{pq}(t, \tau)$ of the proposed large-scale RIS assisted mobile system under dynamic sub-array partition can be re-expressed from (2) to the following expression
\begin{eqnarray}
h_{pq}(t, \tau)   \hspace*{-0.225cm}&=&\hspace*{-0.225cm}  \sqrt{ \Omega_{\text{RIS}}(t) } \sum^{N_{\text{sub}}(t)}_{n_{\text{sub}} = 1} \sum^{M_{\text{sub}}(t)}_{m_{\text{sub}} = 1}   h^{(m_{\text{sub}}, n_{\text{sub}})}_{pq}(t)  \nonumber \\ 
&&\hspace*{-0.225cm}\times \delta\big(\tau - \tau_{(m_{\text{sub}}, n_{\text{sub}})}(t)\big) , 
\end{eqnarray}
where $\tau_{(m_{\text{sub}}, n_{\text{sub}})}(t) = \big( \xi^{T}_{(m_{\text{sub}}, n_{\text{sub}})}(t)  +  \xi^{R}_{(m_{\text{sub}}, n_{\text{sub}})}(t) \big)/c$ is the propagation delay from UAV to MR via the ($m_{\text{sub}}, n_{\text{sub}}$)-th sub-array, and $h^{(m_{\text{sub}}, n_{\text{sub}})}_{pq}(t)$ denotes the CIR from the $p$-th UAV antenna to the $q$-th MR antenna via the ($m_{\text{sub}}, n_{\text{sub}}$)-th sub-array, which can be expressed as
\allowdisplaybreaks[4]
\begin{eqnarray}
h^{(m_{\text{sub}}, n_{\text{sub}})}_{pq}(t)   && \nonumber \\  
&&\hspace*{-2.325cm}=  \sqrt{ \frac{1}{\Upsilon_{pq}(t)} } \sum^{ N^{\text{row}}_{n_{\text{sub}}}(t) }_{n_0 = 1} \sum^{ M^{\text{col}}_{m_{\text{sub}}}(t) }_{ m_0 = 1}   \chi_{mn}(t) e^{j \varphi_{mn}(t)}  \nonumber \\ 
&&\hspace*{-2.1cm}
\times e^{- j \frac{2\pi}{\lambda}  \big( \xi^{T}_{(m_{\text{sub}}, n_{\text{sub}})}(t) + \xi^{R}_{(m_{\text{sub}}, n_{\text{sub}})}(t) \big) }    \nonumber \\  [0.15cm]
&&\hspace*{-2.1cm}\times  e^{ j \frac{2\pi}{\lambda} \big\langle \emph{\textbf{e}}^{T}_{(m_{\text{sub}}, n_{\text{sub}})}(t),  \,  \textbf{d}^{T}_{p}  \big\rangle }       \times       e^{ j \frac{2\pi}{\lambda}  \big\langle \textbf{d}^{m_0 n_0}_{ (m_{\text{sub}}, n_{\text{sub}}) }(t),  \, \emph{\textbf{e}}^{\text{in}}_{(m_{\text{sub}}, n_{\text{sub}})}(t)  \big\rangle }      \nonumber \\  [0.15cm] 
&&\hspace*{-2.1cm}\times  e^{ j \frac{2\pi}{\lambda} \big\langle \emph{\textbf{e}}^{R}_{(m_{\text{sub}}, n_{\text{sub}})}(t),  \, \textbf{d}^{R}_{q}  \big\rangle }      \times       e^{ j \frac{2\pi}{\lambda}  \big\langle \textbf{d}^{m_0 n_0}_{ (m_{\text{sub}}, n_{\text{sub}}) }(t),  \, \emph{\textbf{e}}^{\text{out}}_{(m_{\text{sub}}, n_{\text{sub}})}(t)  \big\rangle }      \nonumber \\ [0.15cm] 
&&\hspace*{-2.1cm}\times  e^{ j \frac{2\pi}{\lambda} \big\langle  \textbf{v}_T t,  \, \emph{\textbf{e}}^{T}_{(m_{\text{sub}}, n_{\text{sub}})}(t)  \big\rangle }      \times       e^{ j \frac{2\pi}{\lambda}  \big\langle  \textbf{v}_R t,  \, \emph{\textbf{e}}^{R}_{(m_{\text{sub}}, n_{\text{sub}})}(t)  \big\rangle }   ,
\end{eqnarray}
in which $n_0 = n - (n_{\text{sub}} - 1)N^{\text{row}}_{n_{\text{sub}} = 1}(t)$ and $m_0 = m - (m_{\text{sub}} - 1)M^{\text{col}}_{m_{\text{sub}} = 1}(t)$, respectively. The $\emph{\textbf{e}}^{T}_{(m_{\text{sub}}, n_{\text{sub}})}(t)$ and $\emph{\textbf{e}}^{R}_{(m_{\text{sub}}, n_{\text{sub}})}(t)$ are the unit directional vectors from UAV and MR to the ($m_{\text{sub}}, n_{\text{sub}}$)-th sub-array, which can be obtained from formula (4) by replacing \{$\alpha^{T}_{mn}(t), \beta^{T}_{mn}(t)$\} and \{$\alpha^{R}_{mn}(t), \beta^{R}_{mn}(t)$\} with \{$\alpha^{T}_{(m_{\text{sub}}, n_{\text{sub}})}(t), \beta^{T}_{(m_{\text{sub}}, n_{\text{sub}})}(t)$\} and \{$\alpha^{R}_{(m_{\text{sub}}, n_{\text{sub}})}(t), \beta^{R}_{(m_{\text{sub}}, n_{\text{sub}})}(t)$\}, respectively. Moreover, the $\textbf{d}^{m_0 n_0}_{ (m_{\text{sub}}, n_{\text{sub}}) }(t)$ represents the distance vector from the center of the ($m_{\text{sub}}, n_{\text{sub}}$)-th sub-array to the ($m_0, n_0$)-th RIS unit, which has the following expression
\begin{eqnarray}
\textbf{d}^{m_0 n_0}_{ (m_{\text{sub}}, n_{\text{sub}}) } (t)  =
\begin{bmatrix}
\hspace*{0.05cm}  \frac{ 2 m_0 - M^{\text{col}}_{m_{\text{sub}}}(t) - 1  }{2} d_c  \hspace*{0.05cm}  \\[0.05cm]
\hspace*{0.05cm}  0   \hspace*{0.05cm}   \\[0.05cm]
\hspace*{0.05cm}  - \frac{ 2 n_0 - N^{\text{row}}_{n_{\text{sub}}}(t) - 1  }{2} d_r   \hspace*{0.05cm}
\end{bmatrix}   ,
\end{eqnarray}
and the $\emph{\textbf{e}}^{\text{in}}_{(m_{\text{sub}}, n_{\text{sub}})}(t)$ and $\emph{\textbf{e}}^{\text{out}}_{(m_{\text{sub}}, n_{\text{sub}})}(t)$ denote the unit direction vectors of the incident and output signals on the plane of the ($m_{\text{sub}}, n_{\text{sub}}$)-th sub-array, respectively, which can be expressed as 
\begin{eqnarray}
\emph{\textbf{e}}^{\text{in}/\text{out}}_{(m_{\text{sub}}, n_{\text{sub}})}(t)  =
\begin{bmatrix}
\hspace*{0.05cm}  \sin\beta^{\text{in}/\text{out}}_{(m_{\text{sub}}, n_{\text{sub}})}(t) \cos\alpha^{\text{in}/\text{out}}_{(m_{\text{sub}}, n_{\text{sub}})}(t)  \hspace*{0.05cm}  \\[0.05cm]
\hspace*{0.05cm}  \cos\beta^{\text{in}/\text{out}}_{(m_{\text{sub}}, n_{\text{sub}})}(t)  \\[0.05cm]
\hspace*{0.05cm}  \sin\beta^{\text{in}/\text{out}}_{(m_{\text{sub}}, n_{\text{sub}})}(t) \sin\alpha^{\text{in}/\text{out}}_{(m_{\text{sub}}, n_{\text{sub}})}(t)   \hspace*{0.05cm}  
\end{bmatrix} ,
\end{eqnarray}
where $\alpha^{\text{in}}_{(m_{\text{sub}}, n_{\text{sub}})}(t)$ and $\beta^{\text{in}}_{(m_{\text{sub}}, n_{\text{sub}})}(t)$ are the azimuth and normal incident angles from the UAV to the ($m_{\text{sub}}, n_{\text{sub}}$)-th sub-array, and $\alpha^{\text{out}}_{(m_{\text{sub}}, n_{\text{sub}})}(t)$ and $\beta^{\text{out}}_{(m_{\text{sub}}, n_{\text{sub}})}(t)$ are the azimuth and normal output angles from the ($m_{\text{sub}}, n_{\text{sub}}$)-th sub-array to the MR, respectively. Their definitions and expressions can be referred to Appendix B in \cite{X_mmWaveUAVRIS} by imposing $\epsilon_I = 0$ as well as replacing \{$x_I; y_I; z_I; \xi^{T}_{\text{RIS}}(t); \xi^{R}_{\text{RIS}}(t)$\} with \{$x_I + A_3 \cos\theta_I; y_I + A_3 \sin\theta_I; z_I - A_4; \xi^{T}_{(m_{\text{sub}}, n_{\text{sub}})}(t); \xi^{R}_{(m_{\text{sub}}, n_{\text{sub}})}(t)$\}, respectively, which are omitted here for brevity.

\begin{figure}[!t]  
\centerline{\includegraphics[width=2.55in,height=1.1in]{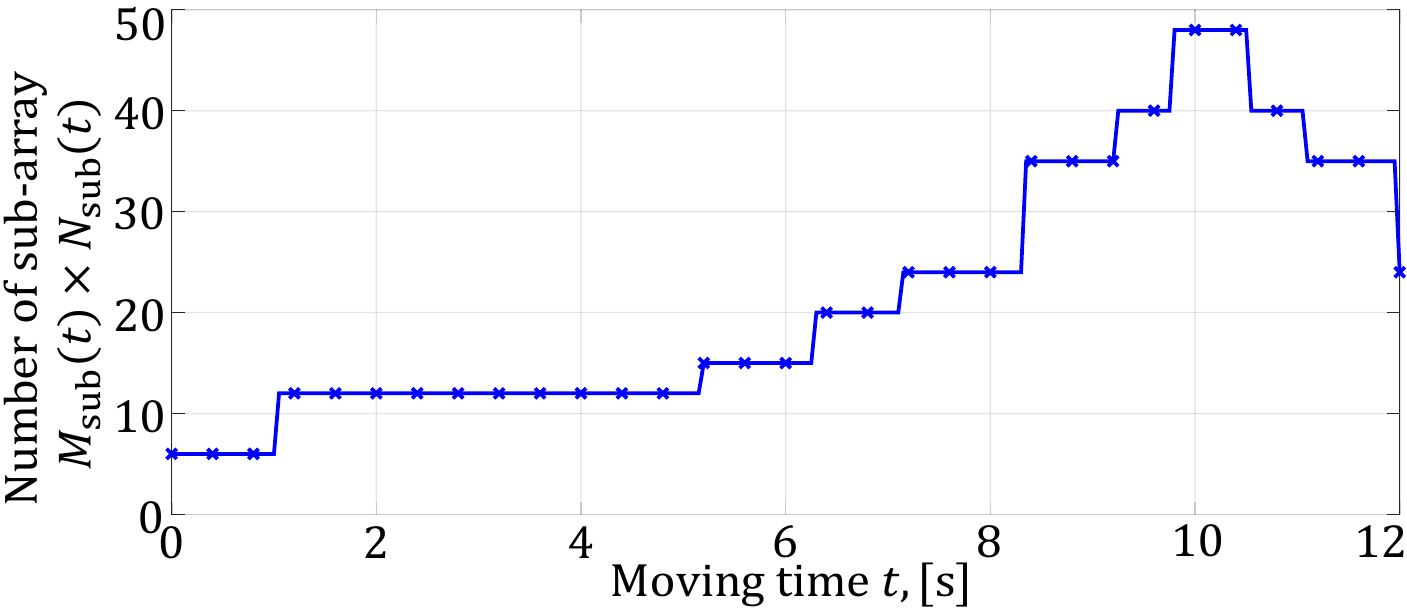}}
\caption{Number of sub-arrays in the proposed large-scale RIS assisted system.}
\end{figure}

Furthermore, the normalized factor $\Upsilon_{pq}(t)$ is expressed as
\begin{eqnarray}
\Upsilon_{pq}(t)  =   \mathbb{E}\Big\{ \Big\vert  \sum^{N}_{n=1}\sum^{M}_{m=1} \chi_{mn}(t) e^{j\big(\varphi_{mn}(t) - \varphi^{dis}_{mn}(t) \big)}   \Big\vert^2 \Big\}  , 
\end{eqnarray}
where $\varphi^{dis}_{mn}(t) =  \frac{2\pi}{\lambda}  \big[ \xi^{T}_{(m_{\text{sub}}, n_{\text{sub}})}(t)  -  \langle \emph{\textbf{e}}^{T}_{(m_{\text{sub}}, n_{\text{sub}})}(t),    \textbf{d}^{T}_{p}  \rangle -  \langle \textbf{d}^{m_0 n_0}_{ (m_{\text{sub}}, n_{\text{sub}}) }(t),  \emph{\textbf{e}}^{\text{in}}_{(m_{\text{sub}}, n_{\text{sub}})}(t)  \rangle  +  \xi^{R}_{(m_{\text{sub}}, n_{\text{sub}})}(t) -  \langle \emph{\textbf{e}}^{R}_{(m_{\text{sub}}, n_{\text{sub}})}(t)$ $, \textbf{d}^{R}_{q} \rangle   -  \langle \textbf{d}^{m_0 n_0}_{ (m_{\text{sub}}, n_{\text{sub}}) }(t),   \emph{\textbf{e}}^{\text{out}}_{(m_{\text{sub}}, n_{\text{sub}})}(t)  \rangle  -  \langle  \textbf{v}_T t,  \, \emph{\textbf{e}}^{T}_{(m_{\text{sub}}, n_{\text{sub}})}(t)  \rangle   -   \langle  \textbf{v}_R t, \emph{\textbf{e}}^{R}_{(m_{\text{sub}}, n_{\text{sub}})}(t) \rangle \big]$ is the distance related phase term via the ($m, n$)-th reflecting unit. And the end-to-end path power gain $\Omega_{\text{RIS}}(t)$ can be expressed as \cite{X_mmWaveUAVRIS}
\begin{eqnarray}
\Omega_{\text{RIS}}(t)   \hspace*{-0.225cm}&=&\hspace*{-0.225cm}   \frac{\lambda^2 d_c d_r}{ (4\pi)^3 }  \times  \mathbb{E}\Big\{ \Big\vert  \sum^{N}_{n=1}\sum^{M}_{m=1}   \sqrt{ \cos\beta^{\text{in}}_{(m_{\text{sub}}, n_{\text{sub}})}(t) }   \nonumber \\ [0.15cm] 
&&\hspace*{-0.225cm}\times  \frac{ \chi_{mn}(t) e^{j\big(\varphi_{mn}(t) - \varphi^{dis}_{mn}(t) \big)} }{ \xi^{T}_{(m_{\text{sub}}, n_{\text{sub}})}(t) \xi^{R}_{(m_{\text{sub}}, n_{\text{sub}})}(t) }    \Big\vert^2 \Big\}  , 
\end{eqnarray}
More specifically, under optimal RIS reflection phase configuration, $\Omega_{\text{RIS}}(t)$ can be rewritten as
\begin{eqnarray}
\Omega^{\text{opt}}_{\text{RIS}}(t)   \hspace*{-0.245cm}&=&\hspace*{-0.245cm}   \frac{\lambda^2 d_c d_r}{ (4\pi)^3 }  \times  \mathbb{E}\Big\{ \Big\vert  \sum^{N_{\text{sub}}(t)}_{n_{\text{sub}} = 1} \sum^{M_{\text{sub}}(t)}_{m_{\text{sub}} = 1}   \sqrt{ \cos\beta^{\text{in}}_{(m_{\text{sub}}, n_{\text{sub}})}(t) }     \nonumber \\ [0.15cm]
&&\hspace*{-0.225cm}\times  \frac{ M^{\text{col}}_{m_{\text{sub}}}(t) N^{\text{row}}_{m_{\text{sub}}}(t) }{\xi^{T}_{(m_{\text{sub}}, n_{\text{sub}})}(t) \xi^{R}_{(m_{\text{sub}}, n_{\text{sub}})}(t) }      \Big\vert^2 \Big\}  .
\end{eqnarray}

\section{Results and Discussions}

The simulation parameters follow the existing work \cite{X_mmWaveUAVRIS} except that a large-scale RIS with dimension $Md_c$ = 3 m and $Nd_r$ = 2 m is considered in this paper. Moreover, we set $M_T$ = $M_R$ = 16, $d_c$ = $d_r$ = $\lambda/3$, $\xi_R$ = 400 m, ($x_I, y_I, z_I$) = (200 m, 50 m, 21 m), $v_R$ = 15 m/s, $\gamma_R$ = 3.075, and RIS with optimal configuration, respectively. The modeling accuracy performance is evaluated by the normalized absolute error $\Delta$, which takes the spherical wavefront model as the~baseline, i.e.,
\begin{eqnarray}
\Delta  =  10\log_{10} \Big\{ \sum^{M_R}_{q = 1} \sum^{M_T}_{p = 1}  \frac{ \vert h_{pq}(t, \tau)  -  h^{\text{spherical}}_{pq}(t, \tau) \vert }{ \vert  h^{\text{spherical}}_{pq}(t, \tau) \vert }  \Big\} .
\end{eqnarray}

Fig. 3 shows the evolution of the number of sub-arrays in the proposed large-scale RIS assisted UAV communication system. It reveals that the deployment of large-scale RIS introduces non-negligible Rayleigh distance, making sub-array partition indispensable even when the terminals remain static. When the terminals move firstly closer to and then farther away from the large-scale RIS, the number of sub-arrays gradually increases to about 50 and then decreases, highlighting the importance of the proposed dynamic sub-array partition scheme.

\begin{figure}[!t]  
\centerline{\includegraphics[width=2.55in,height=1.1in]{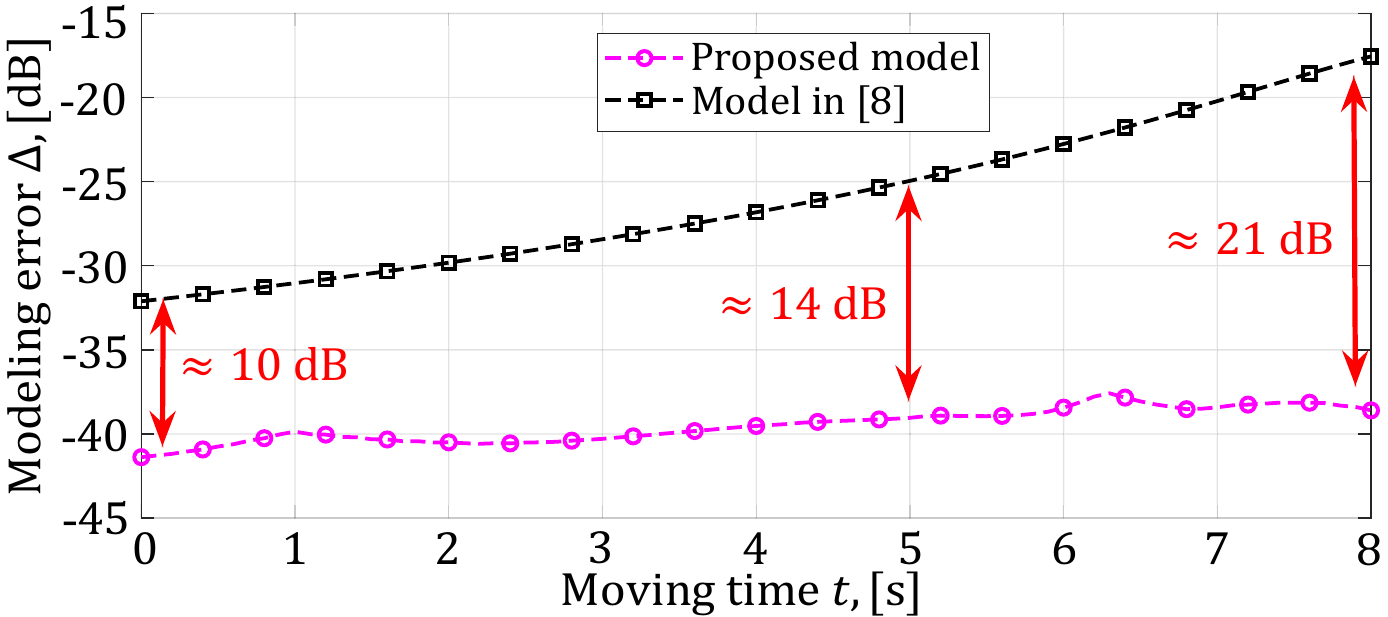}}
\caption{Accuracy performance of the proposed dynamic sub-array partition scheme with respect to moving time, parameter setting follows Fig. 3.}
\end{figure}

\begin{figure}[!t]  
\centerline{\includegraphics[width=2.55in,height=1.1in]{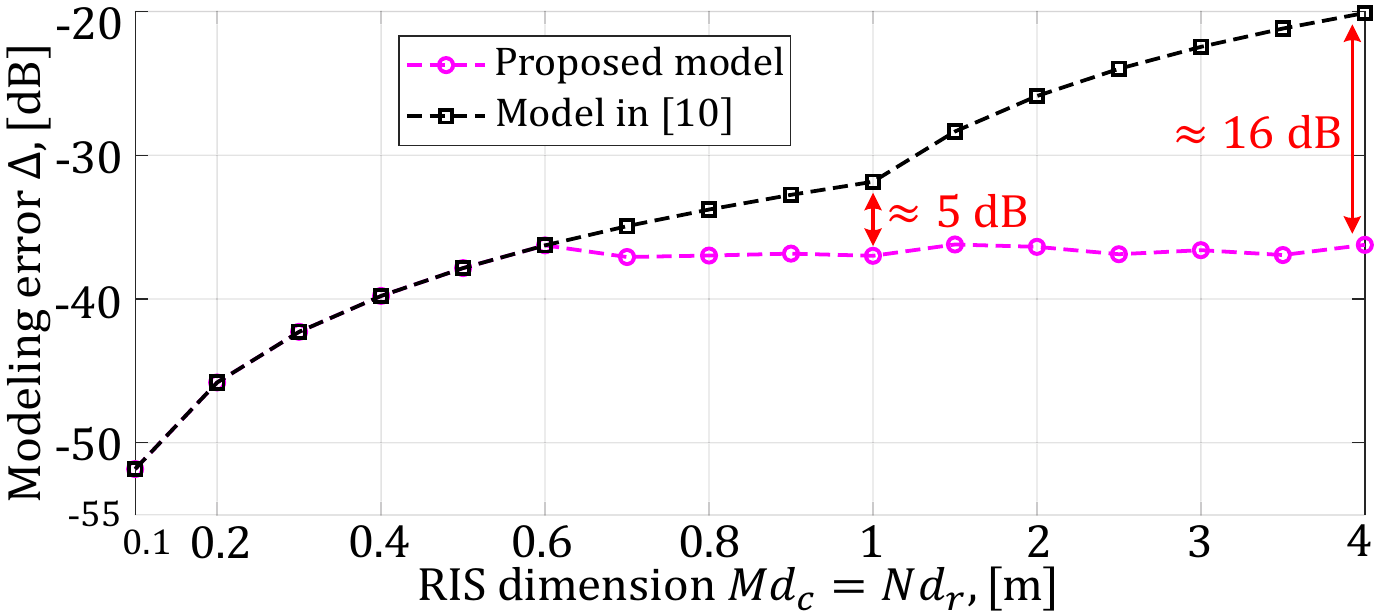}}
\caption{Accuracy performance of the proposed dynamic sub-array partition scheme with respect to RIS dimension, assuming square RIS.}
\end{figure}

We illustrate the modeling accuracy performance of the proposed dynamic sub-array partition scheme in Fig. 4. Under the basic parameter setting, the results indicate that the proposed scheme outperforms the planar wavefront model in \cite{RISsquarelaw} with more than 10 dB accuracy gain. Fig. 4 also reveals that the accuracy performance of the model in \cite{RISsquarelaw} deteriorates obviously as the terminals move closer to the large-scale RIS, whereas the proposed scheme attains almost the same performance. This highlights the advantage of the proposed scheme. Then, Fig. 5 shows the modeling accuracy of the proposed scheme under different RIS dimension when the UAV  and MR moves to (100 m, -50 m, 50 m) and (250 m, 10 m, 0), respectively. It is seen from the figure that the proposed scheme has the same accuracy performance as the planar wavefront model in \cite{Basar_planar} when the RIS dimension is relatively small, i.e., Rayleigh distance smaller than terminal-to-RIS distance, but outperforms the model in \cite{Basar_planar} as the RIS dimension continues to increase. This can be interpreted from the fact that when the RIS dimension is small, no sub-array partition is performed and the proposed model reduces to conventional planar wavefront model, thus attaining the same performance. It is also seen from Figure 5 that the proposed model has the ability to sustain its accuracy performance as the RIS dimension increases, which verifies the effectiveness of the proposed~scheme.

Finally, we compare the complexity as well as accuracy performance among spherical wavefront model, planar wavefront model, and the proposed dynamic sub-array based model for large-scale RIS assisted mobile system in Table II. Specifically, to obtain the CIR, the number of distance and angle parameters required for spherical wavefront model is $7MN$, which are $2MN + 8$ for planar wavefront model and $2MN + 8M_{\text{sub}}(t)N_{\text{sub}}(t)$ for the proposed model, respectively. Since the reflecting units number is much larger than sub-array number in large-scale RIS system, i.e., $MN \gg M_{\text{sub}}(t)N_{\text{sub}}(t)$, the proposed dynamic sub-array based model attains $\sim 70\%$ reduction in computation load with acceptable accuracy. In summary, the proposed model obtains a well balance between modeling complexity and accuracy by performing dynamic sub-array partition, especially for near-field propagation conditions with large dimension RIS.

\begin{table}
\footnotesize  
\centering
\caption{Comparisons Between Models for Large-scale RIS}
\begin{tabular}{|c|c|c|c|}
\hline
\textbf{Model Type}  &\hspace*{-0.375cm} & \textbf{Accuracy}  & \textbf{Computation Load}  \\
\hline
{\makecell[c]{ Spherical Wavefront Model}}  &\hspace*{-0.375cm} &  High    & High   \\
\hline
{\makecell[c]{ Planar Wavefront Model}}   &\hspace*{-0.375cm}  &  Poor  &  Low \\
\hline
{\makecell[c]{ Proposed Model}} &\hspace*{-0.375cm} &  Acceptable   &   Acceptable \\
\hline
\end{tabular}
\vspace*{-0.10cm}
\end{table}

\section{Conclusion}

In this paper, we have considered a mmWave UAV-to-ground mobile communication system with blocking LoS path under the assistance of a large-scale RIS. A dynamic sub-array partition scheme is developed for modeling the proposed large-scale RIS auxiliary system with the purpose of reducing the modeling complexity while sustaining acceptable accuracy performance in a real-time manner, by exploiting the Rayleigh distance criterion as well as the mobile property of the transceivers. The results highlight the importance of dynamic sub-array partition in large-scale RIS assisted mobile networks, and as well reveal the advantages of the proposed scheme in balancing the complexity and accuracy performance as compared to existing models.


\end{document}